\documentclass[aps,pra,twocolumn,superscriptaddress,showpacs,showkeys,amsmath,amssymb]{revtex4}

\usepackage{amsfonts}
\usepackage{amssymb,amsmath}
\usepackage{mathrsfs}
\usepackage{latexsym}
\usepackage{amsmath}
\usepackage[cp1251]{inputenc}
\usepackage{graphicx}
\usepackage{dcolumn}
\usepackage{bm}
\usepackage{color}

\begin{document}
	
	\title{Ground-state properties of dilute one-dimensional Bose gas\\
		 with three-body repulsion}
	
	\author{Volodymyr~Pastukhov\footnote{e-mail: volodyapastukhov@gmail.com}}
	\affiliation{Department for Theoretical Physics, Ivan Franko National University of Lviv,\\ 12 Drahomanov Street, Lviv-5, 79005, Ukraine}

	\date{\today}

	
	
	\begin{abstract}
		We determined perturbatively the low-energy universal thermodynamics of dilute one-dimensional bosons with the three-body repulsive forces. The final results are presented for the limit of vanishing potential range in terms of three-particle scattering length. An analogue of Tan`s energy theorem for considered system is derived in generic case without assuming weakness of the interparticle interaction. We also obtained an exact identity relating the three-body contact to the energy density.
	\end{abstract}
	
	\maketitle
\section{Introduction}
Properties of many-body systems in one spatial dimension due to increasing effect of quantum fluctuations are known to be a very different from their high-dimensional counterparts. The role of statistics in the one-dimensional (1D) geometry attenuates or becomes unimportant at all, when the interaction between particles is switched on. Particularly the bosons fermionize and fermions bosonizide in one dimension. Of course, in order to capture theoretically such a peculiar behavior one requires quite different methods especially since the standard perturbative theory usually fails. And an unique approach that describes the low-energy physics of Bose, Fermi particles and spin systems on equal footing \cite{Haldane,Giamarchi,Cazalilla} is well-elaborated. Another interesting feature of 1D systems is the existence of exact solutions for several non-trivial problems (see \cite{Takahashi} and references therein). But the Bethe ansatz wave function describing properties of these models at low temperatures is adjusted only for a pairwise (typically zero-range) interparticle potential and extensions of this technique on three-, four- and higher-order (effective \cite{Johnson,Hammer,Petrov}) few-body interactions is not found. 

Recently the problem of a three-body interaction in one dimension has attracted much attention in the literature. In particular, conditions for the $N$-body bound state formation in the system of 1D bosons with two-body and three-body attractive interaction was formulated in Ref.~\cite{Nishida}. An explicit analytical expression for the ground and excited energies of trimers composed of bosonic atoms interacting via zero-range two- and three-body forces was derived in \cite{Guijarro} and the three-body pseudo-potential in one dimension was obtained in Ref.~\cite{Pricoupenko_arxiv}.
 
In context of many-body systems, the impact of three-body interaction was recently revealed in the thermodynamics of three-component fermions with attractive three-body potential \cite{Drut} and in the formation of quantum droplet state \cite{Sekino} which stabilizes the 1D bosons with a three-body attraction. In contrast to systems with a continuous translation symmetry, a variety of phases and the quantum critical behavior in the 1D Bose-Hubbard lattice model with a three-body on-site interaction was extensively explored (see for review \cite{Dutta}) in last few years.

The present article deals with the comprehensive study of ground-state properties of the uniform 1D Bose gas with a weak three-body repulsion.

\section{Formulation}
\subsection{Model}
We consider system of $N$ spinless particles of mass $m$ loaded in 1D geometry of volume $L$ with periodic boundary conditions imposed. For simplicity, the interaction between bosons on microscopic level is characterized only by a tree-body potential, although the inclusion of two-body scattering processes is straightforward. The specific form of this three-particle potential is not important for further study except it has to be a short-ranged. Actually the scale associated with the interaction range $r$ is assumed to be much smaller than other length scales, and in order to eliminate the explicit dependence on this parameter and rewrite the final formulae for thermodynamics of 1D Bose gas in terms of universal quantities we adopt the prescription very often referred to Popov. Introducing the auxiliary wave-vector scale $\Lambda<1/r$ that separates the so-called `slowly' and `rapidly' varying fields in the secondly-quantized energy operator we then perform the averaging of the Hamiltonian of `slow' fields over the high-energy states. The resulting effective grand-canonical Hamiltonian governing the dynamics of the system with a three-body interaction at distances larger than $1/\Lambda$ does not depend on range $r$ of the initial three-body potential
\begin{eqnarray}\label{H_eff}
H_{{\rm eff}}=\int dz \,\psi^+(z)\left\{-\frac{\hbar^2}{2m}\partial_{z}^2-\mu
\right\}\psi(z)\nonumber\\
+\frac{1}{3!}\int dz\,\mathcal{T}_3[\psi^+(z)]^3[\psi(z)]^3,
\end{eqnarray} 
where $z\in [0,L]$ is spatial coordinate, $\psi^+(z)$ ($\psi(z)$) is the Bose creation (annihilation) field operator that contains the Fourier harmonics with modulus of the wave-vector less than $\Lambda$ and chemical potential $\mu$ fixes the total number of particles. A strength of the effective local three-body potential in (\ref{H_eff}) is characterized by the cutoff-dependent coupling constant $\mathcal{T}_3$ and has a simple diagrammatic representation (see Fig.~1) 
\begin{figure}[h!]
	\centerline{\includegraphics
		[width=0.09\textwidth,clip,angle=-90]{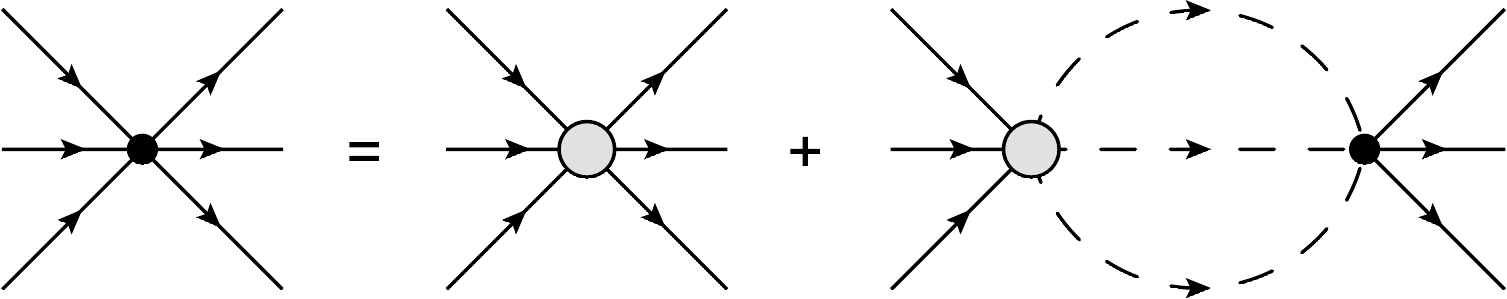}}
	\caption{The effective three-body interaction. The small black dot denotes  $\mathcal{T}_3$, large gray circle stands for the microscopic three-body potential. Dashed lines are the free-particle Green's functions of `rapid' fields (i.e., with $|k|>\Lambda$).}
\end{figure}
\begin{eqnarray}\label{Tau}
\mathcal{T}^{-1}_3=t^{-1}_3-\frac{1}{L^2}\sum_{k, q}\frac{1}{\varepsilon_k+\varepsilon_q+\varepsilon_{k+q}+\epsilon},
\end{eqnarray}
in here $\varepsilon_k=\hbar^2k^2/2m$ is the free-particle dispersion, the summations over wave-vectors are restricted $|k|, |q|<\Lambda$ and we have already assumed that $\mu\ll\hbar^2\Lambda^2/m$ (actually, after elimination of explicit dependence on the interaction range $r$, parameter $\Lambda$ is the largest scale with dimension of inverse length in this problem which should be setted to infinity in final formulae). We also introduced the zero-momentum three-body $t$-matrix $t^{-1}_3=\frac{m}{2\sqrt{3}\pi\hbar^2}\ln\left |\frac{\epsilon_3}{\epsilon}\right|$ which is explicitly related to the three-particle binding energy \cite{Guijarro} $\epsilon_3=-\frac{4e^{-2\gamma}\hbar^2}{ma^2_3}$
($\gamma=0.5772\ldots$ is the Euler-Mascheroni constant and $a_3$ is the three-body scattering length). An auxiliary energy scale $\epsilon>0$ that is of order magnitude of chemical potential $\mu$ for dilute systems and is introduced here only for calculational purposes. A very similar computation scheme was shown \cite{ Konietin,Pastukhov_2D} to be efficient for the perturbative analysis of thermodynamics of 2D Bose gas with a two-body interaction, which technically has much in common with the problem considered here.

It is well-known that there is no condensate in the uniform one-dimensional Bose systems, therefore the use of standard Bogoliubov's technique is questionable in this case, nevertheless leads to correct results \cite{Lieb} in the dilute limit. At same time the ground state of 1D bosons possesses the off-diagonal quasi-long-range order, which can be easily revealed by the hydrodynamic description \cite{Mora_03,Pastukhov_InfraredStr}. In this approach the bosonic fields $\psi^+(z)=\sqrt{n(z)}e^{-i\phi(z)}$, $\psi(z)=e^{i\phi(z)}\sqrt{n(z)}$ are represented in terms of phase $\phi(z)$ and density $n(z)$ operators that satisfy the following commutation relation $[n(z), \phi(z')]=i\delta(z-z')$ (within our discussion this delta-function should be treated as follows: $\delta(z)\to \delta_{\Lambda}(z)=\frac{1}{L}\sum_{|k|<\Lambda}e^{ikz}$).

In terms of phase and density operators the effective Hamiltonian (\ref{H_eff}) reads (argument in fields $n(z)$ and $\phi(z)$ is temporary omitted) 
\begin{align}\label{H_hydro}
H_{{\rm eff}}=\int dz\, \left\{\frac{\hbar^2}{2m}n(\partial_z\phi)^2+\frac{\hbar^2}{8m}\frac{(\partial_z n)^2}{n}-\mu n\right.\nonumber\\
\left.+\frac{1}{3!}\mathcal{T}_3n^3-\frac{1}{2}\mathcal{T}_3\delta_{\Lambda}(0)n^2+\frac{1}{3}\mathcal{T}_3\delta^2_{\Lambda}(0)n\right\}.
\end{align}
The further prescription is well-described in literature: expanding in (\ref{H_hydro}) the density operators near the uniform value $n(z)=n+\delta n(z)$ (where $n=N/L$ and $\int dz\,\delta n(z)=0$) we obtain the Hamiltonian as an infinite series in powers of the density fluctuation $\delta n(z)$. Then, by taking into account only quadratic part of (\ref{H_hydro}) we correctly reproduce the Bogoliubov correction to the ground-state energy of bosons with a three-body repulsive potential, while the higher-order non-linearities can be treated perturbatively in the extremely dilute limit $na_3\ll 1$.

\subsection{Three-body contact}
The presented effective formulation that roughens the length scale and smooths out all details of microscopic three-particle potential, besides its simplicity in calculations of thermodynamic properties of a system, also offers the derivation of Tan`s energy relation for the Bose particles with a three-body repulsion. Particularly by using the effective Hamiltonian (\ref{H_eff}) as well as the Feynman-Hellmann theorem we can evaluate the total energy of the system as sum of mean kinetic $-m\frac{\partial}{\partial m}\langle H_{\rm eff}\rangle$ and potential $t_3\frac{\partial}{\partial t_3}\langle H_{\rm eff}\rangle$ energies
\begin{eqnarray}\label{E_Tan}
\frac{E}{L}=\frac{\hbar^2\mathcal{C}_3}{4\pi m}\ln\left |\frac{\epsilon_3}{\epsilon}\right|+\frac{1}{L}\sum_{k}\varepsilon_k\left\{N_k-\frac{\mathcal{C}_3}{k^2\sqrt{k^2+k^2_{\epsilon}}}\right\},
\end{eqnarray}
where for convenience notation $k^2_{\epsilon}=4m\epsilon/3\hbar^2$ is adopted. The introduced here parameter 
\begin{eqnarray}\label{C_3}
\mathcal{C}_3=\frac{1}{3\sqrt{3}}\left(\frac{m\mathcal{T}_3}{\hbar^2}\right)^2
\langle[\psi^+(z)]^3[\psi(z)]^3\rangle,
\end{eqnarray}
is the 1D three-body analogue of the contact parameter \cite{Tan,Braaten} that determines the leading-order large-$k$ behavior of particle distribution $N_k$. Note that $\mathcal{C}_3$ is finite within the renormalization procedure (\ref{Tau}) in every order of a series expansion over $t_3$ and the ground-state energy determined by Eq.~(\ref{E_Tan}) is independent of auxiliary scale $\epsilon$. Exactly the same situation is realized in 2D systems \cite{Combescot,Valiente} with a pair interparticle interaction.

The knowledge of an exact ultraviolet asymptotics of one-particle momentum distribution $N_k$ allows to obtain the universal short-distance behavior of the one-body density matrix of bosons with a three-body repulsion
\begin{eqnarray}\label{psi_psi_0}
\langle \psi^+(0)\psi(z)\rangle|_{z\to 0}= n+\frac{\mathcal{C}_3}{2\pi}z^2\ln|z|+\ldots
\end{eqnarray}
Recall, that the long-distance properties of $\langle \psi^+(0)\psi(z)\rangle $ are also generic for bosons with an arbitrary type of microscopic repulsive interaction and are fully dictated by the phase fluctuations 
\begin{eqnarray}\label{psi_psi_inf}
\langle \psi^+(0)\psi(z)\rangle|_{z\to \infty}\sim \exp\left\{ -\langle[\phi(z)-\phi(0)]^2\rangle/2\right\}.
\end{eqnarray}
Therefore, by taking into account the exact large-$z$ dependence of correlation function $\langle\phi(z)\phi(0)\rangle$ (which can be most simply derived \cite{Popov_72,Pastukhov_q2D} in the hydrodynamic approach), one easily demonstrates that the one-body density matrix exhibits power-law behavior at large distances
\begin{eqnarray}
\langle \psi^+(0)\psi(z)\rangle|_{z\to \infty}\sim \frac{1}{|z|^{\eta}},
\end{eqnarray}
with the model-dependent exponent $\eta=\frac{1}{2\pi\hbar}\sqrt{\frac{m}{n}\frac{\partial \mu}{\partial n}}$ determined by the compressibility.

\subsection{Dilute limit}
It is usually believed that in a case when interaction between particles is weak the thermodynamic properties of system can be calculated by means of perturbation theory. The latter, of course, does not help to elucidate the accuracy and even convergence of such a series expansion in terms of an arbitrary small parameter ($na_3$ in our case). But a successful application of the second-order perturbation theory \cite{Popov_77} to the exactly solvable Lieb-Liniger model, where it was shown that the approximate series reproduces well the energy of a system up to value of inherent small parameter of order unity, brings some confidence in the perturbative treatment.

We have performed the calculations of the ground-state energy of Bose gas with a three-body short-range repulsion up to the second beyond-mean-field correction. The result of these cumbersome computations (which are closely related to those performed for a system with pairwise interaction \cite{Pastukhov_2D})) is the following 
\begin{eqnarray}\label{E}
E=E_B+\Delta E,
\end{eqnarray}
where an analogue of the Bogoliubov approximation for energy of the system with a three-body potential reads
\begin{eqnarray}\label{E_B}
E_B=\frac{1}{3!}Ln^3\mathcal{T}_3+\frac{1}{2}\sum_{ k}\left(E_k-\varepsilon_k-n^2\mathcal{T}_3\right),
\end{eqnarray}
($E_k=\sqrt{\hbar^4k^4/4m^2+n^2\mathcal{T}_3\hbar^2k^2/m}$ is the Bogoliubov spectrum) and the first-order correction is given by (recall that $|k|,|q|<\Lambda$)
\begin{align}\label{Delta_E}
&\Delta E=-\frac{1}{6N}\sum_{k,q}\frac{1}{\alpha_q\alpha_k\alpha_{k+q}}
\frac{f^2(k,q)}{E_q+E_k+E_{k+q}}\nonumber\\
&+\frac{1}{6N}\sum_{k,q}\left(3\varepsilon_{k+q}/4-n^2\mathcal{T}_3\right)\left(1-\frac{1}{\alpha_k}\right)\nonumber\\
&\times\left(1-\frac{1}{\alpha_q}\right)\left(1-\frac{1}{\alpha_{k+q}}\right)\nonumber\\
&+\frac{1}{2N}\sum_{k,q}n^2\mathcal{T}_3\left(1-\frac{1}{\alpha_k}\right)\left(1-\frac{1}{\alpha_q}\right),
\end{align}
where notations for $\alpha_k=E_k/\varepsilon_k$ and for a symmetric function
\begin{align}\label{f}
f(k,q)=n^2\mathcal{T}_3+\frac{\hbar^2}{4m}\left[kq(\alpha_k-1)(\alpha_q-1)\right.\nonumber\\
\left.-k(k+q)(\alpha_k-1)(\alpha_{k+q}-1)\right.\nonumber\\
\left.-q(k+q)(\alpha_q-1)(\alpha_{k+q}-1)\right],
\end{align}
are used. The above integrals that determine the energy correction and written in the form (\ref{Delta_E}) are logarithmically divergent when $\Lambda \to \infty$. And to obtain some meaningful result we have to replace by using (\ref{Tau}) everywhere in this equation the bare coupling constant $\mathcal{T}_3$ via its expansion over $t_3$. After this, the upper integration limits in $\Delta E$ can be freely lengthened to infinity.
 
We also calculated the three-body contact given by (\ref{C_3}) up to the second beyond-mean-field order. Although any details of these calculations are not presented, but it is easy to verify that product $\mathcal{T}_3^2\langle[\psi^+(z)]^3[\psi(z)]^3\rangle$ remains finite within the above-discussed renormalization procedure. Moreover, the obtained in that way $\mathcal{C}_3$ confirms the exact identity between contact and energy density (a very similar expression is known \cite{Werner} for two-dimensional systems with a pairwise potential)
\begin{eqnarray}\label{E_C_3}
\frac{1}{L}\frac{\partial E}{\partial \ln a_3}=\frac{\hbar^2 \mathcal{C}_3}{2\pi m}.
\end{eqnarray}
This formula along with Tan's energy relation (\ref{E_Tan}) for 1D Bose systems with a three-body interaction constitute the central results of present article.

\section{Perturbative results}
Now we are in position to perform the numerical evaluation of coefficients in the perturbative expansion for energy and three-body contact. The structure of these series is somewhat similar to that for the 2D Bose gas with two-body short-range forces \cite{Schick,Cherny,Mora09,Beane_10}. Actually it is not surprising because of correspondence \cite{Pricoupenko_arxiv} between the two-body problem in 2D and the three-body one in 1D. As a result of this similarity the intrinsic expansion parameter $1/|\ln na_3|$ in our case is also logarithmically dependent on the three-body scattering length $a_3$.

Performing analytical integration over the wave-vector in Eq.~(\ref{E_B}) along with the numerical evaluation of integrals in (\ref{Delta_E}) we find the following small-$na_3$ expansion for energy per particle
\begin{eqnarray}\label{E_res}
\frac{E}{N}=\frac{\pi \hbar^2n^2}{2\sqrt{3}m|\ln na_3|}\left\{1-\frac{4\cdot 3^{1/4}}{\sqrt{\pi|\ln na_3|}}\right.\nonumber\\
\left.-\frac{\frac{1}{2}\ln|\ln na_3|+C_E}{|\ln na_3|}+\ldots \right\},
\end{eqnarray}
where introduce constant reads $C_E=\ln2-\frac{3}{4}\ln3-\frac{1}{2}\ln \pi-\gamma-8.583\ldots =-9.863\ldots$. In Eq.~(\ref{E_res}) the entire Bogoliubov correction and partly $\Delta E$ reproduce the weak-coupling series expansion of the Lieb-Liniger model with the effective two-body potential $\frac{\sqrt{3}\pi\hbar^2n}{m\left|\ln na_3\right|}\delta(z)$, while truly three-body scattering processes contribute only to the third term in brackets. With this result in hands it is easy to evaluate the derivative in Eq.~(\ref{E_C_3}) to obtain the contact for 1D bosons with a three-body repulsion. Although the final expression for $\mathcal{C}_3$ can be explicitly written
\begin{eqnarray}\label{C_res}
\mathcal{C}_3=\frac{\pi^2 n^3}{\sqrt{3}\ln^2na_3}\left\{1-\frac{6\cdot 3^{1/4}}{\sqrt{\pi|\ln na_3|}}\right.\nonumber\\
\left.-\frac{\ln|\ln na_3|+2C_{E}-\frac{1}{2}}{|\ln na_3|}+\ldots \right\},
\end{eqnarray}
the graphical dependence on gas parameter is also plotted for convenience in Fig.~2.
\begin{figure}[h!]
    \centerline{\includegraphics
        [width=0.6
        \textwidth,clip,angle=-0]{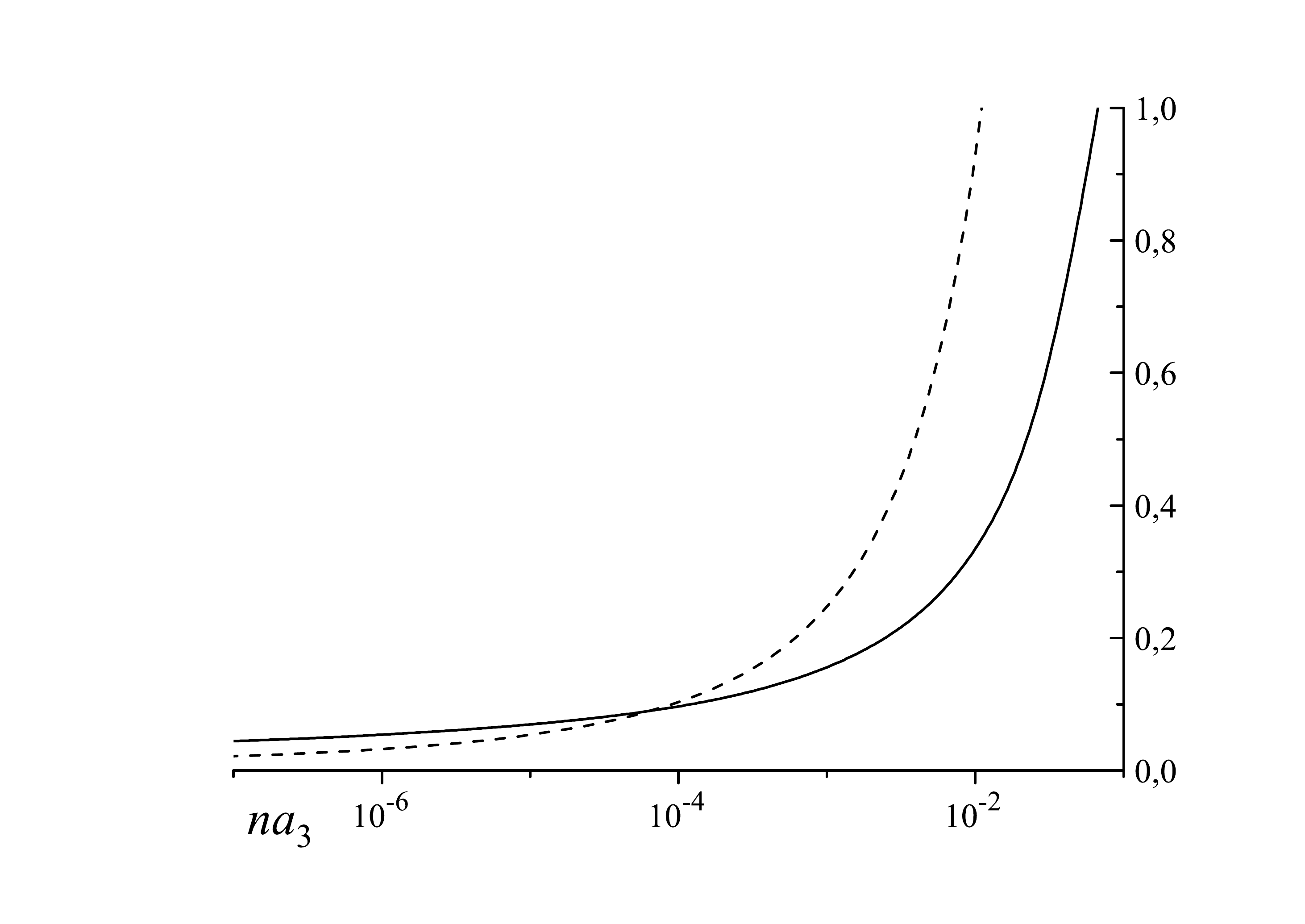}}
    \caption{The energy $E/N$ per particle in units of $\hbar^2n^2/m$ (solid line) and the dimensionless three-body contact $\mathcal{C}_3/n^3$ (dashed line) as functions of gas parameter $na_3$.}
\end{figure}

\section{Conclusions}
In conclusion, we have presented an efficient method that allows to study the universal properties of one-dimensional systems with a short-range three-body interaction. This problem that has a lot in common with the two-dimensional systems with a pairwise interaction was considered by starting from the one-dimensional microscopic model of Bose particles interacting via a finite-range three-body potential. Then applying Popov's original prescription that effectively sums up an infinite series of ladder diagrams we have obtained the Hamiltonian that eliminates the explicit dependence on the range of a three-body potential and correctly describes properties of the system on energy scales per particle much smaller than the recoil energy. It is shown that despite its roughness this effective formulation allows for a very simple derivation of Tan's energy theorem in a case of the Bose gas with a three-body repulsion.

We also performed the perturbative analysis of the ground-state properties of bosons with a triple potential in one dimension. For this purpose we adopted the hydrodynamic approach in terms of density and phase operators to the effective Hamiltonian and working up to the second beyond-mean-field order we calculated the energy of a system and the three-body contact parameter. Based on these calculation an exact identity relating the three-body contact to derivative of the energy density with respect to the three-body scattering length was obtained.

Finally, it should be noted that the presented technique is not restricted to 1D systems and can be easily extended on the description of two- and three-dimensional Bose gases with the short-range three-body potentials.

\begin{center}
	{\bf Acknowledgements}
\end{center}
This work was partly supported by Project FF-30F (No.~0116U001539) from the Ministry of Education and Science of Ukraine.

\newpage

\end{document}